# How to identify the roots of broad research topics and fields? The introduction of RPYS sampling using the example of climate change research

Robin Haunschild*+, Werner Marx*, Andreas Thor**, and Lutz Bornmann***

* Max Planck Institute for Solid State Research
Heisenbergstraße 1, 70569 Stuttgart, Germany
E-mail: w.marx@fkf.mpg.de, r.haunschild@fkf.mpg.de

** University of Applied Sciences for Telecommunications Leipzig (HfTL)
Gustav-Freytag-Str. 43-45, 04277 Leipzig, Germany
E-Mail: thor@hft-leipzig.de

*** Division for Science and Innovation Studies
Administrative Headquarters of the Max Planck Society
Hofgartenstr. 8,
80539 Munich, Germany.
E-mail: bornmann@gv.mpg.de

+ Corresponding author

Abstract

Since the introduction of the reference publication year spectroscopy (RPYS) method and the corresponding program CRExplorer, many studies have been published revealing the historical roots of topics, fields, and researchers. The application of the method was restricted up to now by the available memory of the computer used for running the CRExplorer. Thus, many users could not perform RPYS for broader research fields or topics. In this study, we present various sampling methods to solve this problem: random, systematic, and cluster sampling. We introduce the script language of the CRExplorer which can be used to draw many samples from the population dataset. Based on a large dataset of publications from climate change research, we compare RPYS results using population data with RPYS results using different sampling techniques. From our comparison with the full RPYS (population spectrogram), we conclude that the cluster sampling performs worst and the systematic sampling performs best. The random sampling also performs very well but not as well as the systematic sampling. The study therefore demonstrates the fruitfulness of the sampling approach for applying RPYS.

# Introduction

Thor, Marx, Leydesdorff, and Bornmann (2016) introduced the CRExplorer – a program which can be used to investigate the roots of research fields and topics. For example, the program has been used by Rhaiem and Bornmann (2018) to reveal the historical roots of the new topic in scientometrics of academic efficiency assessments or by Andy Wai Kan (2017) identifying seminal works that built the foundation for functional magnetic resonance imaging studies of taste and food. The CRExplorer facilitates the so-called reference publication year spectroscopy (RPYS) (Marx, Bornmann, Barth, & Leydesdorff, 2014). This statistical method is based on a field- or topic-specific publication set including cited references (CRs). RPYS visualizes CR counts by referenced publication years (RPYs, not to be confused with the method RPYS); years with high counts (especially early years) point to underlying cited publications which might be interpreted as historical roots or landmark papers of a field or topic.

Since the introduction of the RPYS, the method faces the problem of proceeding large datasets which are based on broader topics or fields. The hardware capacities of conventional computers running the CRExplorer are frequently not sufficient enough to process large datasets. To tackle this problem in using the software, we introduce in this paper the technique of drawing several samples from a large dataset and to produce RPYS results based on these samples. The study is based on a large dataset which has been produced by Haunschild, Bornmann, and Marx (2016) to identify the early roots of climate change research (Marx, Haunschild, Thor, & Bornmann, 2017). As we will demonstrate in this study some sampling methods lead to results which are very close to the results from the complete climate change dataset (the population).

By using samples to draw conclusions on populations, the study connects to the recent discussion in the *Journal of Informetrics* around the paper "sampling issues in bibliometric analysis" published by Williams and Bornmann (2016). Both authors demonstrate the relevance of the sampling concept for bibliometric analyses (in the context of inference statistics). Some authors have commented on the paper by questioning the relevance of the sampling topic for the field. In this paper, however, we will demonstrate the fruitfulness of this concept for bibliometric studies.

In the following section "Dataset and Methodology", we describe the climate change dataset which we used in this study to demonstrate the various RPYS sampling methods. The three different sampling methods which are implemented in the CRExplorer are also explained in this section: random, systematic, and cluster sampling. The section "Results" starts with the RPYS based on the complete climate change dataset, i.e. the population dataset (subsection "Population analysis"). The results of the population RPYS constitute the outcome which should be reached by the sampling methods: the closer the RPYS of the sampling method is to the population RPYS, the more appropriate is the method for replacing the population RPYS. The results of the population RPYS revealing the historical roots of climate change research are explained in detail. The subsections "Random sampling", "Systematic sampling", and "Cluster sampling" in the results section present the RPYS results based on the different sampling methods.

All subsections in the "Results" section presenting the RPYS results based on the population and sample data are followed by corresponding subsections, in which the script language of the CRExplorer is explained for performing the specific RPYS. The explanations are provided in detail so that the reader learns how to use the language.

## Dataset and Methodology

### Climate change publications

Our analyses are based on the Web of Science (WoS, Clarivate Analytics) custom data of our in-house database derived from the Science Citation Index Expanded (SCI-E), Social Sciences Citation Index (SSCI), and Arts and Humanities Citation Index (AHCI) produced by Clarivate Analytics (Philadelphia, USA). We used in this study a publication set containing most of the relevant literature regarding climate change research. The set was compiled using a sophisticated method known as "interactive query formulation". A set of key papers was retrieved and a reformulated search query based on the keyword analysis of key papers was constructed (Wacholder, 2011). The search was restricted to the publication years 1980-2014 and to the document types "article" and "review". A detailed description of the search process for retrieving the relevant publications on climate change can be found in Haunschild, et al. (2016).

In total, the publication set (the population) comprises 222,060 publications and 10,932,050 CRs in 4,004,082 distinct CR variants. An earlier RPYS study by Marx, Haunschild, Thor, et al. (2017) has analyzed the RPYs before 1971. The restriction to RPYs before 1971 reduced the number of distinct CR variants to 239,887. This reduction of the number of cited references (NCR) made the RPYS analysis feasible. The CRs published between 1970 and 2014 comprise 6,594,657 CRs in 3,728,879 distinct CR variants. The main memory requirements rise with the number of unique CR variants which makes it impossible to analyze the RPYS using the full climate change dataset on a current standard computer. Thus, the dataset is well suited to demonstrate different sampling methods in this study.

**Sampling methods**

If a dataset contains numerous CRs from many publications, the full dataset cannot be completely imported in the CRExplorer because of restrictions by the available main memory on the computer of many users. To tackle this problem, the user has the option to draw one of the following three types of samples from the full dataset. The samples are based on different methods for selecting a subset from the original set of all CRs (the population) (Levy & Lemeshow, 2008):

(1) **Random Sampling**: The sample of CRs is randomly selected from the population where every possible combination of *n* CRs from the population has the same chance of being selected. For example, if the user wants to import a sample of 100 CRs out of the population of 400 overall CRs, CRExplorer randomly selects 25% of all CRs.

(2) **Systematic sampling**: Systematic sampling is a very popular sampling method (Levy & Lemeshow, 2008) whereby elements are selected from an ordered sampling frame. Here, a given number of CRs is used to select the sample uniformly distributed over the list of all CRs of the citing publications. For example, if the user wants to import 100 CRs out of 400 overall CRs, CRExplorer systematically selects 25% of the list of all CRs by picking the $1^{st}$, $5^{th}$, $9^{th}$, and so on CR.

(3) **Cluster sampling**: Cluster sampling is not a sampling frame which is based on individual units, but on clusters of units. Thus, clusters of units are sampled instead of individual units. The CRExplorer randomly selects one year from the citing publication years which lie between two given years set by the user of the program. Then, all CRs in the papers published in this year are selected as a sample and are imported. The results of Bornmann and Mutz (2015) reveal that the restriction on all CRs from a recent citing year leads to very similar results as the consideration of all CRs from several citing years in references analysis.

## Results

### Population analysis

The results of the RPYS based on the population data which are shown in Figure 1 (the population spectrogram) serve as baseline for the comparison with the results based on the three sampling methods. The figure presents the NCRs for each RPY. Frequently occurring RPYs show up as distinct peaks within the RPYS spectrogram. The highest peak in Figure 1 with the most CRs is visible for RPY = 2000.

For this study, we restrict the RPYS analysis to the RPYs from 1970 to 2010 and use the results for comparison with the RPYS results from various sampling methods. We connect with this focus to the study by Marx, Haunschild, Thor, et al. (2017) who analyzed the very early roots of climate change research. Thus, the results of the RPYS are not only of interest in the comparison of samples and population, but also for revealing landmark publications in climate change research from the past which have been published more recently.

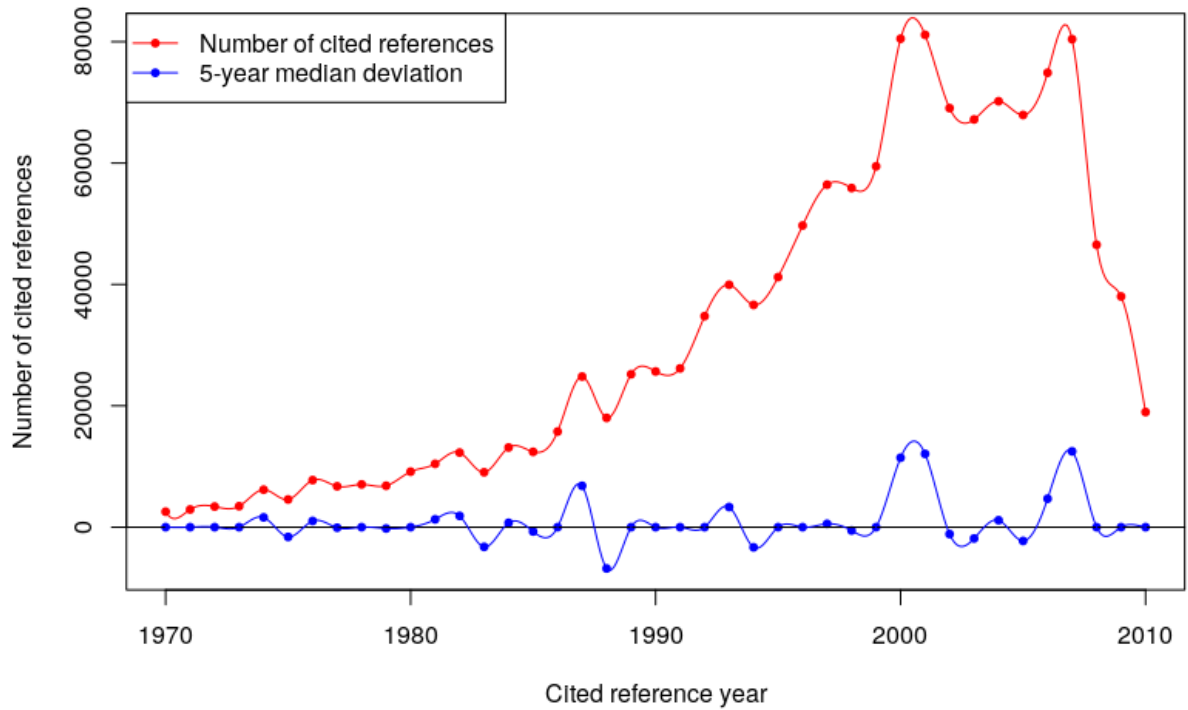

**Figure 1:** Annual distribution of CRs throughout the period 1970-2010 which have been cited in climate change publications (published between 1980 and 2014)

The RPYS in Figure 1 does not only show the NCRs (in red), but also the five-year median deviation (in blue). Thus, the blue line is the deviation of the NCRs in each year from the median for the NCRs in the two previous, the current, and the two following years. This deviation from the five-year median provides a curve smoother than the one in terms of absolute numbers. Using the five-year median deviation curve, peaks in the data can be identified more easily than with the absolute numbers, since each year is compared with its adjacent years. Although we have calculated the RPYS until 2014, we show the spectrogram in Figure 1 only until 2010 to ensure a referencing window of at least three years. The spectrogram features nine more or less

pronounced peaks at the following RPYs: 1974, 1976, 1982, 1984, 1987, 1993, 2000/2001, 2004, and 2007. Table 1 lists the CRs which occur most frequently within the peak RPYs.

**Table 1:** Most frequently CRs, their titles, and NCR values from selected RPYs in Figure 1

| No | RPY | Reference and title | NCR |
|---|---|---|---|
| CR1 | 1974 | AKAIKE H, 1974, IEEE T AUTOMAT CO AC, V19, P716<br>A new look at the statistical model identification | 688 |
| CR2 | 1974 | DEAN WE, 1974, J SEDIMENT PETROL, V44, P242<br>Determination of carbonate and organic matter in calcareous sediments and sedimentary rocks by loss of ignition: comparison with other methods | 527 |
| CR3 | 1974 | ARAKAWA A, 1974, J ATMOS SCI, V31, P674<br>Interaction of a clumulus cloud ensemble with the large-scale environment, part I | 493 |
| CR4 | 1976 | FRITTS HC, 1976, TREE RINGS CLIMATE<br>Book title: Tree rings and climate | 1,515 |
| CR5 | 1976 | HAYS JD, 1976, SCIENCE, V194, P1121<br>Variations in the Earth's orbit: pacemaker of the Ice Ages | 923 |
| CR6 | 1982 | NORTH GR, 1982, MON WEA REV, V110, P699<br>Sampling errors in the estimation of empirical orthogonal functions | 676 |
| CR7 | 1982 | RASMUSSON EM, 1982, MON WEA REV, V110, P354<br>Variations in tropical sea surface temperature and surface wind fields associated with the Southern Oscillation/El Nino | 614 |
| CR8 | 1982 | POST WM, 1982, NATURE, V298, P156<br>Soil carbon pools and world life zones | 542 |
| CR9 | 1984 | WIGLEY TML, 1984, J CLIM APPL METEOROL, V23, P201<br>On the average value of correlated time series, with applications in dendroclimatology and hydrometeorology | 793 |
| CR10 | 1984 | IMBRIE J, 1984, MILANKOVITCH CLIMA 1, P269<br>The orbital theory of Pleistocene climate: support from a revised chronology of the marine $d^{18}O$ record | 768 |
| CR11 | 1987 | ROPELEWSKI CF, 1987, MON WEATHER REV, V115, P1606<br>Global and regional scale precipitation patterns associated with the El Nino/Southern Oscillation | 1,243 |
| CR12 | 1987 | BARNSTON AG, 1987, MON WEATHER REV, V115, P1083<br>Classification, seasonality and persistence of low-frequency atmospheric circulation patterns | 1,067 |
| CR13 | 1987 | MARTINSON DG, 1987, QUATERNARY RES, V27, P1<br>Age dating and the orbital theory of the Ice Ages: development of a high-resolution 0 to 300,000-year chronostratigraphy | 1,047 |
| CR14 | 1993 | STUIVER M, 1993, RADIOCARBON, V35, P215 | 2,332 |
|      |      | Extended $^{14}$C data base and revised Calib 3.0 $^{14}$C age calibration program                                                                                        |       |
| CR15 | 1993 | DANSGAARD W, 1993, NATURE, V364, P218<br>Evidence for general instability of past climate from a 250-kyr ice-core record                                                  | 1,872 |
| CR16 | 2000 | NAKICENOVIC N, 2000, SPECIAL REPORT EMISS<br>Special report on emissions scenarios                                                                                        | 1,470 |
| CR17 | 2000 | GORDON C, 2000, CLIM DYNAM, V16, P147<br>The stimulation of SST, sea ice extents and ocean heat transports in a version of the Hadley Centre coupled model without flux adjustments | 1,283 |
| CR18 | 2001 | HOUGHTON JT, 2001, CLIMATE CHANGE 2001<br>Climate change 2001: the scientific basis                                                                                        | 2,566 |
| CR19 | 2001 | ZACHOS J, 2001, SCIENCE, V292, P686<br>Trends, rhythms, and aberrations in global climate 65 ma to present                                                                | 1,779 |
| CR20 | 2001 | *IPCC, 2001, CLIM CHANG 2001 SCI                                                                                                                                          | 1,625 |
| CR21 | 2004 | THOMAS CD, 2004, NATURE, V427, P145<br>Extinction risk from climate change                                                                                                | 1,765 |
| CR22 | 2004 | REIMER PJ, 2004, RADIOCARBON, V46, P1029<br>Intcal04 terrestrial radiocarbon age calibration, 0-26 cal kyr bp                                                              | 1,225 |
| CR23 | 2007 | SOLOMON S, 2007, CLIM CHANG 2007, P19<br>IPCC Fourth Assessment Report: Climate Change 2007 (AR4), Working Group I Report "The Physical Science Basis", Technical Summary | 4,125 |
| CR24 | 2007 | *IPCC, 2007, CLIM CHANG 2007 PHYS<br>IPCC Fourth Assessment Report: Climate Change 2007 (AR4), Working Group I Report "The Physical Science Basis"                        | 2,622 |
| CR25 | 2007 | MEEHL GA, 2007, CLIM CHANG 2007, P747<br>IPCC Fourth Assessment Report: Climate Change 2007 (AR4), Working Group I Report "The Physical Science Basis", Chapter 10: Global climate projections.. | 1,882 |
| CR26 | 2007 | *IPCC, 2007, CLIMATE CHANGE 2007<br>IPCC Fourth Assessment Report: Climate Change 2007 (AR4)                                                                              | 1,743 |

The 26 CRs in Table 1 can be categorized into four different groups of climate change research papers. Nine CRs (CR4, CR5, CR9, CR10, CR13, CR14, CR15, CR19, and CR22) can be assigned to the disciplines paleoclimatology and dating techniques. The corresponding papers deal with the orbital theory of the Ice Ages, the instability of the climate of the past, and dendrochronology in connection with climate research. Six CRs (CR3, CR6, CR7, CR11, CR12, and CR17) are concerned with meteorology. The publications mainly present measured data or modelling results with regard to the atmospheric and oceanic circulation systems. These two sets of CRs are distributed more or less equally over the selected time span. Since the year 2000, however, IPCC reports increasingly appear as the most-frequently CRs. Seven CRs (CR16, CR18, CR20, and CR23-CR26) are part of IPCC reports, mostly related to the scientific basis of climate change and emission scenarios of greenhouse gases. Finally, there are four CRs (CR1, CR2, CR8, and CR21) which deal with various other issues in climate change research, e.g. biological and statistical studies about effects from climate change.

We use the spectrogram in Figure 1 and the most frequently cited publications in Table 1 to judge the reliability of the different sampling methods results which are presented in following sections.

**Using the script language for the population analysis**

We employed the script language of the CRExplorer to produce the results in Figure 1 and Table 1. The language can be applied instead of using the menus of the graphical user interface of CRExplorer. A separate JAR file is necessary to use the language (this file can be downloaded from http://www1.hft-leipzig.de/thor/crexplorer/CitedReferencesExplorerScript.jar). We started by analyzing the CRs in all climate change papers on a machine with 512 GB of main memory (RAM, random access memory). The CRE and CSV files which are necessary for a RPYS analysis of all CRs published between 1970 and 2014 can be produced using the following CRExplorer script:

```
set(n_pct_range: 0, median_range: 2)

importFile(file: "savedrecs.txt", type: "WOS", RPY: [1970, 2014, false], PY: [1980, 2014, false], maxCR: 0)

info()

cluster(threshold: 0.75, volume: true, page: true, DOI: false)

merge()

removeCR(N_CR: [0, 99])

saveFile(file: "savedrecs.cre")

exportFile(file: "savedrecs_CR.csv", type: "CSV_CR")

exportFile(file: "savedrecs_GRAPH.csv", type: "CSV_GRAPH")
```

**Listing 1:** CRExplorer script to analyze the CRs in the WoS file savedrecs.txt

Listing 1 imports the WoS file with the complete climate change data. Furthermore, it identifies variants of the same CR in the dataset, cluster them, and merge their occurrences (NCRs) (Thor, et al., 2016). Three export files are saved in different formats.

The set function in the listing can be used to change options of the settings dialog in the CRExplorer. We set usage of two neighboring RPYs for calculation of the median deviation in this case, i.e. a five-year median deviation. The option n_pct_range: 0 is set here and in the following scripts for purely technical reasons. This option does not change the results presented in this study.

The function importFile is needed to import WoS or Scopus files. We supply options to restrict the CRs to RPYs between 1970 and 2014 and publication years of citing publications between 1980 and 2014. The value of maxCR can be used to limit the number of imported CRs. A value of 0 means no limit. The function info prints a brief line of information to the screen.

With the function cluster, we clustered the imported CRs automatically by using a similarity threshold of 0.75 considering volume and page. The function merge merges the clustered CR variants. Consistent with Marx, Haunschild, Thor, et al. (2017), we removed all CR variants occurring less than 100 times with the removeCR function.

The functions saveFile and exportFile allow us to save the results of our analysis in different formats: the CRE-internal file format, the list of CRs in CSV file format, and the data to produce the RPYS graph in CSV format (see Figure 1). The latter can be used to produce RPYS graphs with the Stata command plotrpys (https://ideas.repec.org/c/boc/bocode/s458378.html) and the R package BibPlots (https://cran.r-project.org/web/packages/BibPlots/index.html). 35 GB RAM are needed to cluster the CRs of RPY 1970-2014.

**Random sampling**

In an attempt to cover a range from small to large number of samples, we performed seven different random sample RPYS analyses using 10, 50, 100, 500, 1,000, 2,500, and 10,000 samples with 50,000 CRs in each sample. Figure 2 shows the results of the merged samples in comparison with the population spectrogram (full RPYS). As the samples are of different size, they had to be scaled. We used $f = max(\text{NCR}_{sample,RPY})/max(\text{NCR}_{full, RPY})$ as a scaling factor. The samples do not fully reproduce the population spectrogram but most of the relevant peaks also occur in all of the samples. It seems that a few (10 or 50) random samples are sufficient to obtain a first impression of the RPYS.

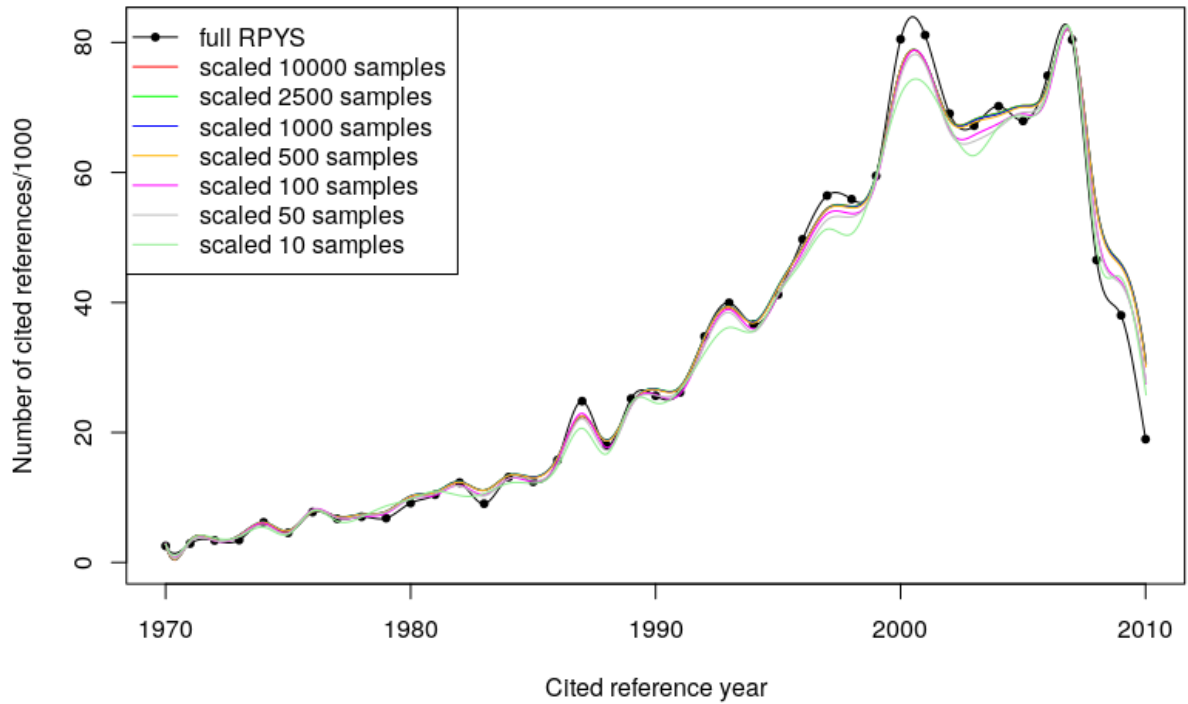

**Figure 2:** Annual distribution of random samples of the CRs throughout the period 1970-2010 which have been cited in climate change publications (published between 1980 and 2014)

The differences between the samples can be seen more clearly in Figure 3 where the difference between each sample and the RPYS with 10,000 samples is shown. The random sampling seems to converge rather slowly with the sample size, but the RPYS with 500 samples seems to be a good compromise between accuracy and computational time. Each sample needed approximately one minute of computational time on our Intel® Xeon® E5-2640 with 2.6GHz so that 500 samples can be calculated within a day or overnight. 10,000 samples of 50,000 CRs each needed about a week on the same PC. Due to the slow convergence of the random sampling, we present the most important references under the peaks for the results from 10,000 samples in Table 2.

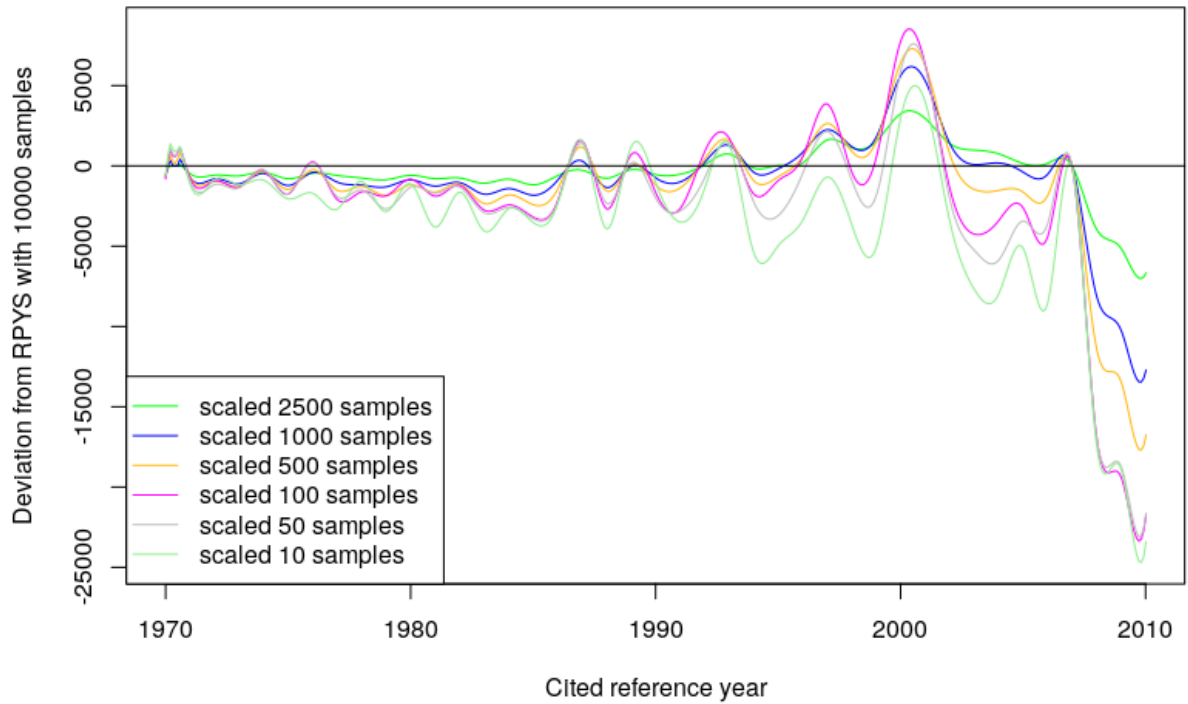

**Figure 3:** Deviation of the randomly sampled RPYS results from the RPYS based on 10,000 samples

**Table 2:** Most frequently CRs from selected RPYs with their NCR values using 10,000 random samples

| No | RPY | Reference | NCR |
|---|---|---|---|
| CR1 | 1974 | AKAIKE H, 1974, IEEE T AUTOMAT CO AC, V19, P716 | 676 |
| CR2 | 1974 | DEAN WE, 1974, J SEDIMENT PETROL, V44, P242 | 509 |
| CR3 | 1974 | ARAKAWA A, 1974, J ATMOS SCI, V31, P674 | 459 |
| CR4 | 1976 | FRITTS HC, 1976, TREE RINGS CLIMATE | 1,515 |
| CR5 | 1976 | HAYS JD, 1976, SCIENCE, V194, P1121 | 913 |
| CR6 | 1982 | NORTH GR, 1982, MON WEA REV, V110, P699 | 649 |
| CR7 | 1982 | RASMUSSON EM, 1982, MON WEA REV, V110, P354 | 606 |
| CR8 | 1982 | POST WM, 1982, NATURE, V298, P156 | 538 |
| CR9 | 1984 | WIGLEY TML, 1984, J CLIM APPL METEOROL, V23, P201 | 789 |
| CR10 | 1984 | IMBRIE J, 1984, MILANKOVITCH CLIMA 1, P269 | 768 |

| CR11 | 1987 | MARTINSON DG, 1987, QUATERNARY RES, V27, P1 | 1,042 |
| CR12 | 1987 | ROPELEWSKI CF, 1987, MON WEATHER REV, V115, P1606 | 1,008 |
| CR13 | 1987 | BARNSTON AG, 1987, MON WEATHER REV, V115, P1083 | 832 |
| CR14 | 1993 | DANSGAARD W, 1993, NATURE, V364, P218 | 1,854 |
| CR15 | 1993 | STUIVER M, 1993, RADIOCARBON, V35, P215 | 1,559 |
| CR16 | 2000 | NAKICENOVIC N, 2000, SPECIAL REPORT EMISS | 1,470 |
| CR17 | 2000 | GORDON C, 2000, CLIM DYNAM, V16, P147 | 1,274 |
| CR18 | 2001 | HOUGHTON JT, 2001, CLIMATE CHANGE 2001 | 2,566 |
| CR19 | 2001 | ZACHOS J, 2001, SCIENCE, V292, P686 | 1,707 |
| CR20 | 2001 | *IPCC, 2001, CLIM CHANG 2001 SCI | 1,625 |
| CR21 | 2004 | THOMAS CD, 2004, NATURE, V427, P145 | 1,746 |
| CR22 | 2004 | REIMER PJ, 2004, RADIOCARBON, V46, P1029 | 1,175 |
| CR23 | 2007 | SOLOMON S, 2007, CLIM CHANG 2007, P19 | 4,125 |
| CR24 | 2007 | *IPCC, 2007, CLIM CHANG 2007 PHYS | 2,622 |
| CR25 | 2007 | MEEHL GA, 2007, CLIM CHANG 2007, P747 | 1,882 |
| CR26 | 2007 | *IPCC, 2007, CLIMATE CHANGE 2007 | 1,743 |

A comparison of the CRs in Table 2 with those in Table 1 shows that the same CRs occur as relevant peak papers in the sampling procedure as well as in the population analysis. However, the order of the peak papers is different for RPYs 1987 and 1993. In the case of eight CRs (CR4, CR10, CR17. CR19, and CR23-CR26), even the NCR value of the sampling result agrees with the result from the population RPYS analysis.

**Systematic sampling**

Analogously to the random sampling, we calculated seven different RPYSs of different sample sizes. The scaled sampling RPYS results and the population spectrogram are shown in Figure 4. Essentially all peaks except the small peak in 2004 are reproduced by all samples. Also in the case of the systematic sampling, a small sample size seems to be enough to resemble the most important features of the population spectrogram.

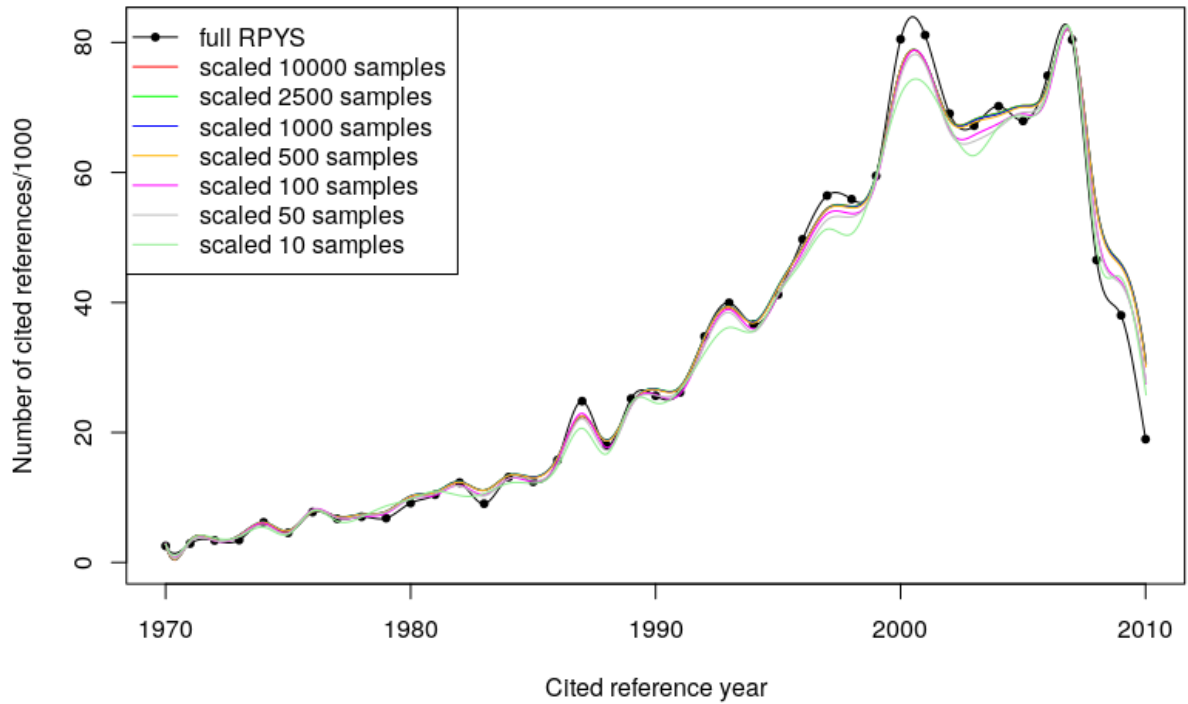

**Figure 4:** Annual distribution of systematic samples of the CRs throughout the period 1970-2010 which have been cited in climate change publications (published between 1980 and 2014)

The differences between the RPYS with 10,000 samples and the RPYS results with smaller sample sizes are displayed in Figure 5. In the case of the climate change literature, the systematic sampling converges faster than the random sampling. The difference between the RPYS result of 500 samples and larger samples seems to be insignificant. However, smaller sample sizes do not seem to be sufficient to resemble the RPYS accurately.

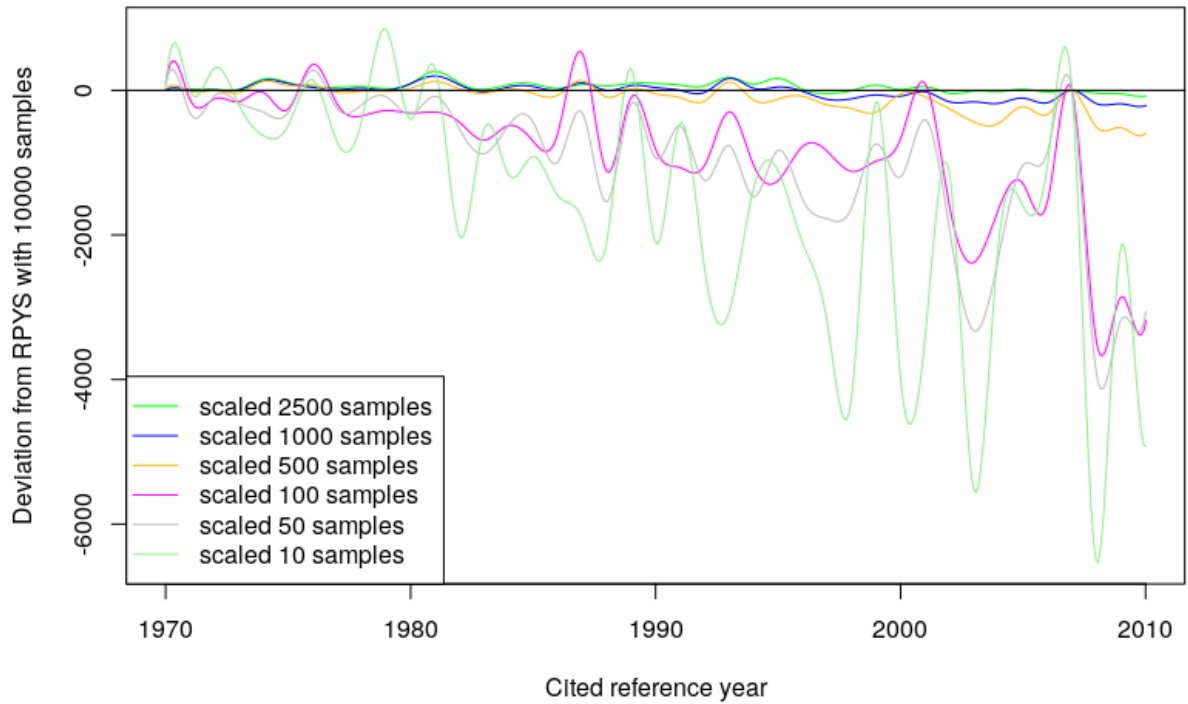

**Figure 5:** Deviation of the systematically sampled RPYS results from the RPYS based on 10,000 samples

The most frequently occurring CRs according to 500 systematic samples for the peak RPYs are shown in Table 3.

**Table 3:** Most frequently CRs from selected RPYs with their NCR values using 500 systematic samples

| No | RPY | Reference | NCR |
|---|---|---|---|
| CR1 | 1974 | AKAIKE H, 1974, IEEE T AUTOMAT CO AC, V19, P716 | 522 |
| CR2 | 1974 | DEAN WE, 1974, J SEDIMENT PETROL, V44, P242 | 482 |
| CR3 | 1974 | ARAKAWA A, 1974, J ATMOS SCI, V31, P674 | 427 |
| CR4 | 1976 | FRITTS HC, 1976, TREE RINGS CLIMATE | 1,510 |
| CR5 | 1976 | HAYS JD, 1976, SCIENCE, V194, P1121 | 896 |

| CR6  | 1982 | NORTH GR, 1982, MON WEA REV, V110, P699           | 620   |
| CR7  | 1982 | RASMUSSON EM, 1982, MON WEA REV, V110, P354       | 579   |
| CR8  | 1982 | POST WM, 1982, NATURE, V298, P156                 | 510   |
| CR9  | 1984 | WIGLEY TML, 1984, J CLIM APPL METEOROL, V23, P201 | 766   |
| CR10 | 1984 | IMBRIE J, 1984, MILANKOVITCH CLIMA 1, P269        | 747   |
| CR11 | 1987 | MARTINSON DG, 1987, QUATERNARY RES, V27, P1       | 1,035 |
| CR12 | 1987 | ROPELEWSKI CF, 1987, MON WEATHER REV, V115, P1606 | 1,000 |
| CR13 | 1987 | BARNSTON AG, 1987, MON WEATHER REV, V115, P1083   | 816   |
| CR14 | 1993 | DANSGAARD W, 1993, NATURE, V364, P218             | 1,851 |
| CR15 | 1993 | STUIVER M, 1993, RADIOCARBON, V35, P215           | 1,552 |
| CR16 | 2000 | NAKICENOVIC N, 2000, SPECIAL REPORT EMISS         | 1,459 |
| CR17 | 2000 | GORDON C, 2000, CLIM DYNAM, V16, P147             | 1,268 |
| CR18 | 2001 | HOUGHTON JT, 2001, CLIMATE CHANGE 2001            | 2,551 |
| CR19 | 2001 | ZACHOS J, 2001, SCIENCE, V292, P686               | 1,696 |
| CR20 | 2001 | *IPCC, 2001, CLIM CHANG 2001 SCI                  | 1,619 |
| CR21 | 2004 | THOMAS CD, 2004, NATURE, V427, P145               | 1,735 |
| CR22 | 2004 | REIMER PJ, 2004, RADIOCARBON, V46, P1029          | 1,165 |
| CR23 | 2007 | SOLOMON S, 2007, CLIM CHANG 2007, P19             | 4,109 |
| CR24 | 2007 | *IPCC, 2007, CLIM CHANG 2007 PHYS                 | 2,614 |
| CR25 | 2007 | MEEHL GA, 2007, CLIM CHANG 2007, P747             | 1,875 |
| CR26 | 2007 | *IPCC, 2007, CLIMATE CHANGE 2007                  | 1,737 |

A comparison of Table 1 and Table 3 shows that all top papers of the population RPYS also appear as top papers in the RPYS from 500 systematic samples. Only the order of the top papers is different for RPYs 1987 and 1993. The ordering of the top papers is the same as in the population RPYS for all other RPYs. Even the NCR agrees quite well in most cases. Mainly, the reference Stuiver M, 1993, is significantly underestimated in terms of the NCRs. It seems from our results that the systematic sampling with 500 samples each can be used to approximate the population spectrogram very well.

## Using the script language for random and systematic sampling

The script language can be extended using the Java program language. Every user can expand the capabilities of the CRExplorer by writing such extensions. One CRExplorer extension is available at https://github.com/andreas-thor/cre/blob/master/crs/packages/Loop.crs: Loop.crs.

This extension simplifies loop programming in the CRExplorer script language. The analysis via sampling procedures was made using the extension Loop.crs. In this case, ten random samples of 50,000 CRs were drawn from the population of CRs. They were clustered and merged. Afterwards, CRs referenced only once were removed.

```
use("Loop.crs").with {
    forEachUnion(count:10, dir: "TMPDIR_with_full_path", {index ->
        set(n_pct_range: 0, median_range: 2)
        importFile(file: "savedrecs.txt", type: "WOS", RPY: [1970, 2014, false], PY: [1980, 2014, false], sampling: "RANDOM", maxCR: 50000, offset: index+1)
        info()
        cluster(threshold: 0.75, volume: true, page: true, DOI: false)
        merge()
        removeCR(N_CR: [0, 1])
    })
    saveFile(file: "savedrecs_rs_10.cre")
    exportFile(file: "savedrecs_rs_10_CR.csv", type: "CSV_CR")
    exportFile(file: "savedrecs_rs_10_GRAPH.csv", type: "CSV_GRAPH")
}
```

**Listing 2:** CRExplorer script to analyze 10 random samples of 50,000 CRs from the WoS file savedrecs.txt

Most functions from Listing 2 were already explained in the comments regarding Listing 1. The extension Loop.crs provides the functions forEachUnion and forEach. Both functions provide loops. The number of cycles is provided as the value of count (here 10). The functions differ in their behavior after the loops are finished. forEach performs no further action whereas forEachUnion merges the CRE files of each cycle to a final CRE data set. The parameter dir can be provided but is optional. If parameter dir is not provided, the system default temporary directory is used. If there is too few disc space, the CRExplorer stops with an error message. Furthermore, if dir is provided, the temporary files of each cycle are kept and can be used later on using other CRExplorer script files. The variable index is available in the loop and runs from 0 to count-1. The importFile function contains two additional arguments compared to Listing 1. The parameter sampling can be set to "RANDOM" (as in this example) or "SYSTEMATIC". Two of the sampling methods can be selected this way. The argument offset: index+1 instructs the CRExplorer to skip the first index+1 CRs. This is not necessary for the random sampling, but very important for the systematic sampling. The systematic sampling uses an equidistant set of CRs from the data file. Without the offset option, all samples would contain the same CRs.

The argument maxCR: 50000 restricts the sample size to 50,000 CRs which easily fit into 1 GB RAM, although about 250,000 CRs could be imported per GB from the climate change publication set. However, merging of the samples needed more RAM depending on the number of samples. As multiple samples need more memory than single samples, we deem it appropriate to restrict the sample sizes in our study consistently to 50,000 CRs per sample.

We conducted a series of merging tests determining the number of samples we were able to merge with a certain amount of RAM. The results are shown in Table 4. However, the number of samples and the amount of RAM should be seen as guiding values as they may differ between publication set types and sampling methods. Especially, the values obtained for the random sampling of course strongly depend on the random samples drawn.

**Table 4:** Amount of RAM necessary to merge a certain number of samples with 50,000 CRs each

| Amount of RAM | Number of merged systematic samples | Number of merged random samples |
|---|---|---|
| 1 GB | 29 | 134 |
| 2 GB | 64 | 450 |
| 4 GB | 148 | 4,664 |
| 6 GB | 296 | 9,521 |
| 8 GB | > 500 | >10,000 |

Suppose the user has less than 8 GB of RAM available but still would like to merge 500 systematic samples of 50,000 CRs each, one can also merge in batches, e. g., merging four batches of 125 samples each is possible with 4 GB RAM. However, the resulting CR variants might differ somewhat as they might be determined differently in the various merging steps. In the case of cluster sampling, 2 GB were enough to analyze the publication year 2011 and it was possible to process the publication year 2014 with 4 GB RAM.

The function removeCR in Listing 2 now contains a lower threshold than in the case of the population spectrogram. We propose to use the following rule of thumb for calculating the number of CRs to be removed:

$$NCR_{\text{threshold}}(\text{sample}) = round\left(\frac{NCR_{\text{threshold}}(\text{full})}{NCR_{\text{full}} / NCR_{\text{sample}}}\right)$$

(1)

The number of CRs of each sample ($NCR_{\text{sample}}$) and of the population ($NCR_{\text{full}}$) can be determined via the function analyzeFile. The syntax of analyzeFile is analogous to the one of importFile. This rule of thumb results in our current case in:

$$NCR_{\text{threshold}}(\text{sample}) = round\left(\frac{100}{6{,}594{,}657/50{,}000}\right) \approx round(0.758) = 1$$

(2)

**Cluster sampling**

For cluster sampling, the CRExplorer randomly selects one year from the given set of citing publication years. Then, all CRs from the papers in this year are selected and imported. As an exploration of the cluster sampling, we used the publication years 2011, 2012, 2013, and 2014 and compared the corresponding spectrograms with the population spectrogram (see Figure 6). All cluster sample spectrograms in the figure resemble only the peak in 2007 quite well. No other peak is reproduced properly. The shoulder in 2009 is much too pronounced in the sample RPYS results in comparison with the population spectrogram.

It seems from these results that the cluster sampling should not be recommended for RPYS. It should be explored in future studies, whether the cluster sampling approach is appropriate for other publication sets. We could imagine, for instance, that this approach is feasible for research topics which have been started only a few years ago. In these cases, the CRs in the single citing years might be so uniform that the cluster sampling could work.

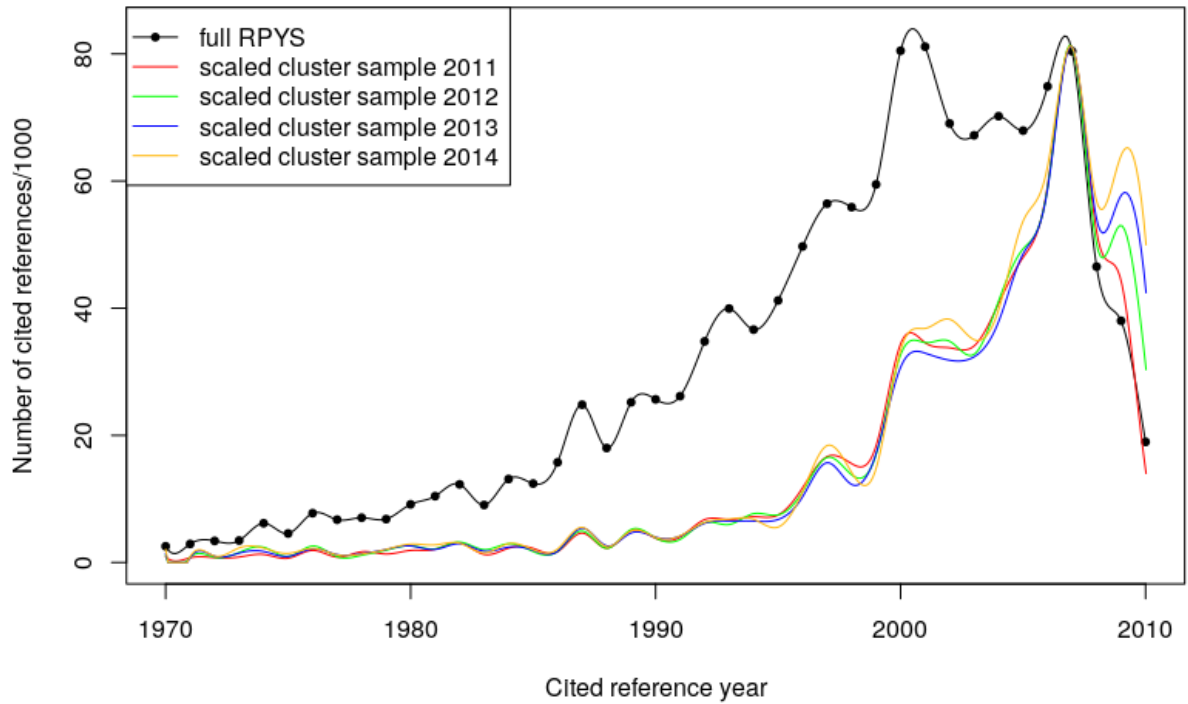

**Figure 6:** Annual distribution of cluster samples of the CRs throughout the period 1970-2010 which have been cited in climate change publications (published between 2011 and 2014)

## Using the script language for cluster sampling

The cluster sampling was performed using CRExplorer scripts like the one in Listing 3. First, the CRs of the citing year 2011 were imported into the CRExplorer. Second, the CRs were clustered using volume and page number but not DOI. Third, the equivalent CRs were merged. Finally, CRs which are referenced 15 times or less were removed from the data set, and the resulting CRE and CSV files were saved.

```
        set(n_pct_range: 0, median_range: 2)

        importFile(file: " savedrecs.txt", type: "WOS", RPY: [1970, 2014, false], PY:
        [2011, 2011, false])

        cluster(threshold: 0.75, volume: true, page: true, DOI: false)

        merge()

        removeCR(N_CR: [0, 15])

        saveFile(file: "savedrecs_cs_2011.cre")

        exportFile(file: "savedrecs_ cs_2011_CR.csv", type: "CSV_CR")

        exportFile(file: "savedrecs_ cs_2011_GRAPH.csv", type: "CSV_GRAPH")
```

**Listing 3:** CRExplorer script to perform a cluster sample from the WoS file savedrecs.txt

In this study, we selected the publication year specifically in order to use the appropriate number of CRs to be removed after merging according to our proposed rule of thumb (see section "Using the script language for random and systematic sampling"). Alternatively, one can use PY: [1980, 2014, false], sampling: "CLUSTER" in Listing 3 to randomly select a citing year. We prefer for this study to select the citing year to systematically determine the threshold for removal of CRs. We used removeCR(N_CR: [0, 15]) for 2011, removeCR(N_CR: [0, 18]) for 2012, removeCR(N_CR: [0, 22]) for 2013, and removeCR(N_CR: [0, 24]) for 2014 in accordance with Eq. (1).

## Discussion and Conclusions

Since the introduction of the RPYS method (and the corresponding program CRExplorer), many studies have been published revealing the historical roots of topics, fields, and researchers (see, e.g., Barth, Marx, Bornmann, & Mutz, 2014; Leydesdorff, Bornmann, Comins, Marx, & Thor, 2016; Leydesdorff, Bornmann, Marx, & Milojevic, 2014; Marx, et al., 2014; Marx, Haunschild,

French, & Bornmann, 2017; Wray & Bornmann, 2014). The application of the method was restricted up to now by the available memory of the computer used for running the CRExplorer. This meant that many users could not perform RPYS for broader research fields or topics. In this study, we present various sampling methods to solve this problem. The study therefore demonstrates the fruitfulness of the sampling approach for bibliometric studies. Some comments following the paper by Williams and Bornmann (2016) questioned the usefulness of this approach for bibliometric studies.

The statistical analysis of large datasets with the CRExplorer becomes more prevalent, since it has become possible with the new program version to import data from CrossRef (see https://www.crossref.org). The user of CrossRef gains free access to meta-data of publications which can be (1) downloaded as files and imported in the CRExplorer or (2) directly imported by using the CRExplorer search interface for CrossRef data. Especially the use of the search interface allows fast access on comprehensive CR data from publications.

In this study, we introduce the script language of the CRExplorer which can be used to draw many samples from the population dataset (see also the handbook of the program at www.crexplorer.net). The language can be applied instead of using the menus in the program. Script languages are standard in statistical software to automate the process of empirical analysis. Once a script has been produced for a given dataset, the script can be used for further similar datasets. Scripts fulfill an important function in the replicability and reproducibility of empirical studies. Are script, dataset, and program for a published study available, the results in the manuscript can be reproduced (and possible errors identified). Although replicability and reproducibility are essential components of the open science movement (Cumming & Calin-Jageman, 2016), scripts are scarcely available for popular bibliometric software, such as VOSviewer or CitNetExplorer. Thus, the user of the CRExplorer script language receives an impression, how the script language of bibliometric software could be designed.

Based on a large dataset of publications from climate change research, we compare RPYS results using population data with RPYS results using sampling data. We show RPYS results for three different sampling techniques: random sampling, systematic sampling, and cluster sampling. From our comparison with the full RPYS (population spectrogram), we conclude that the cluster sampling performs worst and the systematic sampling performs best. The random sampling also performs very well but not as well as the systematic sampling. Merging 500 systematic samples of 50,000 CRs each reproduces the population RPYS rather accurately and also the same peak CRs are found in the sampled spectrogram as in the population spectrogram. Merging 10,000 random samples also results in the same peak CRs as obtained from the population RPYS results.

It is unknown if our findings can be transferred to other research fields than climate change. Studying different publication sets might make it necessary to increase the sample sizes or the number of samples drawn, or it might be possible to obtain good RPYS results with smaller sample sizes or number of samples. We would like to encourage other studies to check which sample sizes and number of samples are needed to approximate the population spectrogram accurately enough.

## Acknowledgements

The bibliometric data used in this paper are from an in-house database developed and maintained by the Max Planck Digital Library (MPDL, Munich) and derived from the Science Citation Index Expanded (SCI-E), Social Sciences Citation Index (SSCI), Arts and Humanities Citation Index (AHCI) provided by Clarivate Analytics (Philadelphia, Pennsylvania, USA).

# References

Andy Wai Kan, Y. (2017). Identification of seminal works that built the foundation for functional magnetic resonance imaging studies of taste and food. *Current Science, 113*(7), 1225-1227.

Barth, A., Marx, W., Bornmann, L., & Mutz, R. (2014). On the origins and the historical roots of the Higgs boson research from a bibliometric perspective. *The European Physical Journal Plus, 129*(6), 1-13. doi: 10.1140/epjp/i2014-14111-6.

Bornmann, L., & Mutz, R. (2015). Growth rates of modern science: A bibliometric analysis based on the number of publications and cited references. *Journal of the Association for Information Science and Technology, 66*(11), 2215-2222. doi: 10.1002/asi.23329.

Cumming, G., & Calin-Jageman, R. (2016). *Introduction to the New Statistics: Estimation, Open Science, and Beyond*: Taylor & Francis.

Haunschild, R., Bornmann, L., & Marx, W. (2016). Climate Change Research in View of Bibliometrics. *Plos One, 11*(7), 19. doi: 10.1371/journal.pone.0160393.

Levy, P. S., & Lemeshow, S. (2008). *Sampling of Populations: Methods and Applications*. Hoboken, NJ, USA: Wiley.

Leydesdorff, L., Bornmann, L., Comins, J., Marx, W., & Thor, A. (2016). Referenced Publication Year Spectrography (RPYS) and Algorithmic Historiography: The Bibliometric Reconstruction of András Schumbert's Œuvre. In W. Glänzel & B. Schlemmer (Eds.), *András Schubert-A World of Models and Metrics. Festschrift for András Schubert's 70th birthday* (pp. 79-96): International Society for Scientometrics and Informetrics.

Leydesdorff, L., Bornmann, L., Marx, W., & Milojevic, S. (2014). Referenced Publication Years Spectroscopy applied to iMetrics: *Scientometrics*, *Journal of Informetrics*, and a relevant subset of JASIST. *Journal of Informetrics, 8*(1), 162-174. doi: 10.1016/j.joi.2013.11.006.

Marx, W., Bornmann, L., Barth, A., & Leydesdorff, L. (2014). Detecting the historical roots of research fields by reference publication year spectroscopy (RPYS). *Journal of the Association for Information Science and Technology, 65*(4), 751-764. doi: 10.1002/asi.23089.

Marx, W., Haunschild, R., French, B., & Bornmann, L. (2017). Slow reception and under-citedness in climate change research: A case study of Charles David Keeling, discoverer of the risk of global warming. *Scientometrics, 112*(2), 1079-1092. doi: 10.1007/s11192-017-2405-z.

Marx, W., Haunschild, R., Thor, A., & Bornmann, L. (2017). Which early works are cited most frequently in climate change research literature? A bibliometric approach based on Reference Publication Year Spectroscopy. *Scientometrics, 110*(1), 335-353. doi: 10.1007/s11192-016-2177-x.

Rhaiem, M., & Bornmann, L. (2018). Reference Publication Year Spectroscopy (RPYS) with publications in the area of academic efficiency studies: what are the historical roots of this research topic? *Applied Economics, 50*(13), 1442-1453. doi: 10.1080/00036846.2017.1363865.

Thor, A., Marx, W., Leydesdorff, L., & Bornmann, L. (2016). Introducing CitedReferencesExplorer (CRExplorer): A program for Reference Publication Year Spectroscopy with Cited References Standardization. *Journal of Informetrics, 10*(2), 503-515.

Wacholder, N. (2011). Interactive Query Formulation. *Annual Review of Information Science and Technology, 45*, 157-196. doi: 10.1002/aris.2011.1440450111.

Williams, R., & Bornmann, L. (2016). Sampling issues in bibliometric analysis. *Journal of Informetrics, 10*(4), 1253-1257.

Wray, K. B., & Bornmann, L. (2014). Philosophy of science viewed through the lense of "Referenced Publication Years Spectroscopy" (RPYS). *Scientometrics*, 1-10. doi: 10.1007/s11192-014-1465-6.